\documentclass[aps,showpacs,preprintnumbers,eqsecnum,amsmath,amssymb]{revtex4}

\usepackage{float}
\usepackage{graphics,epsfig}
\usepackage{graphicx}
\usepackage{dcolumn}
\usepackage{bm}

\begin{document}

\title{ Late-time evolution of charged massive Dirac fields
in the Kerr-Newman background}
\author{Xi He}\thanks{ Email:jin$_{-}$hexi@126.com}\ \ \ \
\author{Jiliang Jing}\thanks{Corresponding author, Email:
jljing@hunnu.edu.cn}
\affiliation{ Institute of Physics and  Department of Physics, \\
Hunan Normal University,\\ Changsha, Hunan 410081, P. R. China }

\vspace*{0.2cm}
\begin{abstract}
We investigate both the intermediate late-time tail and the
asymptotic tail behavior of the charged massive Dirac fields in the
background of the Kerr-Newman black hole. We find that the
intermediate late-time behavior of charged massive Dirac fields is
dominated by an inverse power-law decaying tail without any
oscillation, which is different from the oscillatory decaying tails
of the scalar field. We note that the dumping exponent depends not
only on the angular quantum numbers $m$, the separation constant
$\lambda$ and the rotating parameter $a$, but also on the product
$seQ$ of the spin weight of the Dirac field and the charges of the
black hole and the fields. We also find that the decay rate of the
asymptotically late-time tail is $t^{-5/6}$, and the oscillation of
the tail has the period of $2\pi / \mu$ which is modulated by two
types of long-term phase shifts.
\end{abstract}

%\keywords{Black hole, Dirac field, Late-time evolution}

\pacs{04.30.-w, 04.62.+v, 97.60.Lf.}

\maketitle

\section{Introduction}

The evolution of field perturbation around a black hole consists
roughly of three stages \cite{Frolov98}. The first one is an
initial wave burst coming directly from the source of perturbation
and is dependent on the initial form of the original field
perturbation. The second one involves the damped oscillations
called the quasinormal modes, the frequencies and damping times of
which are entirely fixed by the structure of the background. The
last one is a power-law tail \cite{tail1} \cite{tail2} behavior of
the waves at very late time, which is caused by backscattering of
the gravitational field.

The late-time evolution of various field perturbations outside a
black hole has important implications for two major aspects of black
hole physics: the no-hair theorem and the mass-inflation
scenario\cite{Poisson}\cite{Burko}. Therefore, since Wheeler
\cite{w1}\cite{w2} introduced the no-hair theorem which states that
black holes are completely characterized only by three externally
observable parameters: mass, electrical charge, and angular
momentum, the decay rate of the various fields have been extensively
studied \cite{eg.1}-\cite{6}. Price and Burko \cite{tail1} studied
the massless neutral external perturbations and found that the
late-time behavior for a fixed ~$r$ is dominated by the factor
$t^{-(2l+3)}$ for each multiple moment. In Refs. \cite{7}\cite{8}
the massless late-time tail for the gravitational, electromagnetic,
neutrino and scalar perturbations had also been considered in the
case of the Kerr black hole. Starobinskii, Novikov
\cite{Starobinskii}  and Burko \cite{L.M.Burko} analyzed the
evolution of a massive scalar field in the Reissner-Norstr\"om
background, and they found that, because of the mass term, there are
poles in the complex plane closer to the real axis than in the
massless case, which leads to inverse power-law behavior with
smaller indices than the massless case. Hod and Piran \cite{13}
pointed out that, if the field mass $\mu$ is small, namely ~$\mu
M\ll 1 $, the oscillatory inverse power-law behavior $ \Phi\sim
t^{-(l+3/2)}\sin(\mu t)$ dominates as the intermediate late-time
tails in the Reissner-Nordstr\"{o}m background. Xue and Wang
\cite{Wang} studied the massive scalar wave propagation in the
background of a Reissner-Nordstr\"m black hole by using numerical
simulations and found that the relaxation process depends only on
the field parameter and does not depend on the spacetime parameter
for $\mu M\ll 1$. The very late-time tails of the massive scalar
fields in the Schwarzschild and Reissner-Nordstr\"{o}m background
were  studied in Refs. \cite{14,15} and in the Kerr case was first
investigated numerically in Ref. \cite{4418}, and it has been
pointed out that the oscillatory inverse power-law behavior of the
dominant asymptotic tail is approximately given by
~$t^{-5/6}\sin(\mu t)$ which is slower than the intermediate ones.
Jing \cite{eg.7} studied the late-time tail behavior of massive
Dirac fields in the Schwarzschild black-hole geometry and found that
the asymptotic behavior of the massive Dirac fields is dominated by
a decaying tail without any oscillation in which the dumping
exponent depends not only on the multiple number of the wave mode
but also on the mass of the Dirac field, and also noted that the
decay of the massive Dirac field is slower than that of the massive
scalar field. Recently, late-time tails of self-interacting
(massive) scalar fields in a dilaton spacetime and in higher
dimensional black hole were investigated by Moderski and Rogatko
\cite{Moderski 1}\cite{ Moderski 2} , and they found that the
intermediate asymptotic behavior of the self-interacting scalar
field is determined by an oscillatory inverse power-law decaying
tail. Konoplya and Molina \cite{ Konoplya} studied the late-time
behavior of massive vector field in the background of Schwarzschild
black holes, and found three functions with different decay law
depending on the multiple number $l$ at intermediately late times
and a asymptotically late-time decay law $\propto t^{-5/6}sin(\mu
t)$ which was independent of $l$.

Although much attention has been paid to the study of the
late-time behaviors of the neutral scalar, gravitational,
electromagnetic in static and stationary black-hole backgrounds,
however, to my best knowledge, at the moment only the late-time
evolution of the charged massless scalar field was investigated by
Hod and Piran \cite{4}-\cite{6} and their conclusion was that a
charged scalar hair outside a charged black hole is dominated by a
~$t^{-(2l+2)}$ tail which decays slower than a neutral one. Jing
\cite{eg.8} recently studied the late-time evolution of charged
massive Dirac fields in the background of a Reissner-Norstr\"om
black hole note that the dumping exponent also depends on the
product of the spin weight of the Dirac field and the charges of
the black hole and the Dirac fields. The main purpose of this
paper is extend the study of  the late-time tail evolution of the
charged massive Dirac fields to the stationary Kerr-Newman
black-hole background.

The organization of this paper is  as follows. In Sec.II the
decoupled charged  massive Dirac equations in the Kerr-Newman
spacetime are presented by using Newman-Penrose formalism. In
Sec.III the black-hole Green's function is introduced and the
late-time evolution of the charged massive Dirac fields in the
Kerr-Newman background is investigated. Sec.IV is devoted to a
summary. In appendix we calculate the separation constant
$\lambda^2$ of the charged massive Dirac equation.

\section{Charged massive Dirac equations in the Kerr-Newman spacetime}

In a curve spacetime the Dirac equations coupled to a electromagnetic
field can be expressed
as \cite{as}
\begin{eqnarray} \label{metric}
&&\sqrt{2}  (\nabla_{A\dot{B}}+ie A_{A \dot{B}}) P^{A}+i
\mu \bar{Q}_{\dot{B}}=0,\nonumber \\
&&\sqrt{2}  (\nabla_{A\dot{B}}-ieA_{A\dot{B}}) Q^{A}+i \mu
\bar{P}_{\dot{B}}=0,
\end{eqnarray}
where $\nabla_{A\dot{B}}$ represents the covariant differentiation,
$P^{A}$ and $Q^{A}$ are the two-component spinors representing the
wave function, $\bar{P}_{\dot{B}}$ is the complex conjugate of
$P_{B}$, $A_{A\dot{B}}$ is a four-vector describes the
electromagnetic field, and $\mu$ and $e$ are the mass and charge of
the Dirac particle. Plank units are used, so $\hbar=c=G=1$.

Using the Newman-Penrose formalism \cite{formalism} with a
null tetrad $(\vec{l},\vec{n},\vec{m},\vec{\bar{m}})$,
 the equation (\ref{metric})
becomes
\begin{eqnarray}\label{NP}
&&(D+\varepsilon-\rho+ieA_{0\dot{0}}) P^{0}+(\bar{\delta}
+\pi-\alpha+ieA_{1\dot{0}}) P^{1}={1\over \sqrt{2}}  i \mu
\bar{Q}^{\dot{1}},\nonumber
\\&&(\Delta+\mu-\gamma+ieA_{1\dot{1}}) P^{1}+(\delta+
\beta-\tau+ieA_{0\dot{1}}) P^{0}=-{1\over \sqrt{2}} i
\mu\bar{Q}^{\dot{0}},\nonumber \\
&&(D+\bar{\varepsilon}-\bar{\rho}+ieA_{\dot{0}0})
\bar{Q}^{\dot{0}}+(\delta+\bar{\pi}-\bar{\alpha}+
ieA_{\dot{1}0})\bar{Q}^{\dot{1}}=-{1\over
\sqrt{2}} i \mu P^{1},\nonumber \\
&&(\bar{\delta}+\bar{\beta}-\bar{\tau}+ieA_{\dot{0}1})
\bar{Q}^{\dot{0}}+(\Delta+\bar{\mu}-\bar{\gamma}+ieA_{\dot{1}1})
\bar{Q}^{\dot{1}}={1\over \sqrt{2}} i \mu P^{0},
\end{eqnarray}
where
\begin{eqnarray}
&&A_{0\dot{0}}=\overrightarrow{A} \cdot
{\overrightarrow{l}}=-{Q r\over \Delta},\nonumber \\
&&A_{0\dot{1}}=\overrightarrow{A} \cdot
{\overrightarrow{m}}=A_{1\dot{0}}=\overrightarrow{A}
\cdot {\overrightarrow{\bar{m}}}=0, \nonumber \\
&&A_{1\dot{1}}=\overrightarrow{A} \cdot {\overrightarrow{n}}=-{Q
r\over {2(r^{2}+a^{2}cos^{2})}}.
\end{eqnarray}

For the Kerr-Newman spacetime, the null tetrad can be taken as
\begin{eqnarray}
&&l^{\mu}=({r^2+a^2\over \Delta},~1,~0,~{a\over \Delta}),\nonumber \\
&&n^{\mu}={1\over 2\rho^{2}}((r^2+a^2),~-\Delta,~0,~a),\nonumber \\
&&m^{\mu}={1\over \sqrt{2}\bar{\rho}}(i a sin\theta,~0,~1,~ {i\over
sin\theta}),
\end{eqnarray}
with
\begin{eqnarray}
\Delta=r^2+a^2+Q^2-2M r,~~~~ \bar{\rho}=r+i a cos\theta,
\end{eqnarray}
where $M$, $Q$ and $a$ represent the mass, charge and angular
momentum per unit mass of the Kerr-Newman black hole respectively.

If we take
\begin{eqnarray}
&&P^{0}={1\over {\bar{\rho}^{\ast}}}e^{-i \omega t+i m
\phi}R_{-{1\over
2}}(r)S_{-{1\over2}}(\theta),\nonumber \\
&&P^{1}=e^{-i \omega t+i m \phi}R_{+{1\over2}}(r)S_{+{1\over
2}}(\theta),\nonumber \\
&&\bar{Q}^{\dot{1}}=e^{-i \omega t+i m
\phi}R_{+{1\over2}}(r)S_{-{1\over 2}}(\theta),\nonumber \\
&&\bar{Q}^{\dot{0}}=-{1\over\bar{\rho}} e^{-i \omega t+i m
\phi}R_{-{1\over2}}(r)S_{+{1\over 2}}(\theta),
\end{eqnarray}
where $\omega$ and $m$ are the energy and angular momentum of the
Dirac particle, after tedious calculation, the equations (\ref{NP}) can be
simplified as
\begin{eqnarray}\label{Sep}
&&{\mathcal{D}}_{0}R_{-{1\over 2}}S_{-{1\over 2}}+{1\over
\sqrt{2}}{\mathcal{L}}_{1\over2}R_{+{1\over
2}}S_{+{1\over2}}={1\over\sqrt{2}}i \mu(r-ia
cos\theta)R_{+{1\over2}}S_{-{1\over2}}, \nonumber   \\
&&\Delta {{\mathcal{D}}^{\dagger}}_{1\over 2}R_{+{1\over
2}}S_{+{1\over 2}}-\sqrt{2}{{\mathcal{L}}^{\dagger}}_{1\over
2}R_{-{1\over 2}}S_{-{1\over 2}}=-\sqrt{2} i \mu (r-i a
cos\theta)R_{-{1\over 2}}S_{+{1\over 2}}, \nonumber   \\
&&{\mathcal{D}}_{0}R_{-{1\over 2}}S_{+{1\over
2}}-{1\over\sqrt{2}}{{\mathcal{L}}^{\dagger}}_{1\over
2}R_{+{1\over 2}}S_{-{1\over 2}}={1\over
\sqrt{2}}i \mu (r+i a cos\theta)R_{+{1\over 2}}S_{+{1\over 2}},\nonumber   \\
&&\Delta{{\mathcal{D}}^{\dagger}}_{1\over 2}R_{+{1\over
2}}S_{-{1\over 2}}+\sqrt{2}{{\mathcal{L}}_{1\over 2}}{R_{-{1\over
2}}S_{+{1\over 2}}}=-\sqrt{2}i\mu(r+iacos\theta)R_{-{1\over
2}}S_{-{1\over 2}},
\end{eqnarray}
with
\begin{eqnarray}
&&{\mathcal{D}}_{n}={\partial\over {\partial}{r}}-{iK\over
\Delta}+2n{r-M\over \Delta},\nonumber \\
&&{{\mathcal{D}}^{\dagger}}_{n}={\partial\over {\partial}{r}}+{i
K\over
\Delta}+2n{r-M\over \Delta},\nonumber \\
&&{\mathcal{L}}_{n}={\partial\over {\partial}{\theta}}-H+n cot\theta,\nonumber \\
&&{{\mathcal{L}}^{\dagger}}_{n}={\partial\over
{\partial}{\theta}}+H+n cot\theta,\nonumber \\
&&K={(r^2+a^2)\omega}- m a - e Q r,\nonumber \\
&&H=a\omega sin\theta-{m\over sin\theta}.
\end{eqnarray}
The equation (\ref{Sep}) can be reduced to the following radial and
angular parts
\begin{eqnarray}\label{SR1}
&&{\mathcal{D}}_{0}{R_{-{1\over 2}}}=(\lambda +i \mu
r){R_{+{1\over 2}}},\\ \label{SR2}
&&\Delta{{{\mathcal{D}}^{\dagger}}_{1\over 2}}{R_{+{1\over
2}}}=(\lambda -i \mu r){R_{-{1\over 2}}},\\ \label{Ang1}
&&{{\mathcal{L}}_{1\over 2}}{S_{+{1\over 2}}}=(-\lambda +\mu a
con\theta)S_{-{1\over 2}},\\ \label{Ang2}
&&{{\mathcal{L}^{\dagger}}_{1\over 2}}{S_{-{1\over
2}}}=(\lambda+\mu a cos\theta)S_{+{1\over 2}},
\end{eqnarray}
where $\lambda$ is a separation constant. Then we can eliminate
$S_{+{1\over2}}$ (or $S_{-{1\over2}}$) from Eqs. (\ref{Ang1}) and
(\ref{Ang2}), and obtain the angular equation
\begin{eqnarray}\label{angular}
{1\over sin\theta}{d\over{d\theta}}\left(sin\theta{d{S}_{s}\over
d\theta}\right)+{{a\mu sin\theta}\over {-2 s \lambda+ a \mu
cos\theta}}{d{S_{s}}\over d\theta}+ \left[ {\mu a sin\theta\over
{\lambda-2s \mu a cos\theta}}H - 2s{\partial H\over
 {\partial\theta}}\right. \nonumber \\
\left.- {1\over 2{sin^2\theta}}+ {\mu a cos\theta\over { 2(-2
s\lambda+a\mu cos\theta) } } -H^2+{1\over 4}{cot^2\theta}+\lambda^2-
\mu^2a^2{cos^2\theta} \right ]S_{s}=0,
\end{eqnarray}
For the slow rotating black hole $\lambda^2$ can be expressed as
(see Appendix A for detail)
\begin{eqnarray}
\lambda^2 &=&\lambda_0^2+\lambda_1^2 a+\lambda_2^2 a^2 +...,
\end{eqnarray}
where $\lambda_0^2$,  $\lambda_1^2$ and $\lambda_2^2$ are
described by Eqs. (\ref{Lambda0}), (\ref{Lambda1}) and
(\ref{Lambda2}).

We can also eliminate $R_{-1/2}$ (or $R_{+1/2}$) from Eqs.
(\ref{SR1}) and (\ref{SR2}), and obtain a radial decoupled equation
for $R_{+1/2}$ (or $R_{-1/2}$). We find both the decoupled equations
for $R_{+1/2}$ and $R_{-1/2}$ can be casted into a single equation
\begin{eqnarray}\label{RR1}
&&\Delta^{-s}{d\over dr}\left(\Delta^{1+s}{dR_{s}\over
dr}\right)+{2 i s \mu \Delta\over {\lambda-2 i s \mu r}}{dR_{s}\over
dr}+P_{s}R_{s}=0,~~~~~~~~~(s=\pm\frac{1}{2})
\end{eqnarray}
with
\begin{eqnarray}
P_{s}&=&{K^2-2 i s K (r-M) \over \Delta}+4 i s\omega r-2 i s e Q+\left(s+\frac{1}{2}\right)\nonumber \\
&&+{i \mu (s+1/2) (r-M)-\mu
K\over {\lambda-2 i s \mu r}}-\mu^2 r^2-\lambda^2.
\end{eqnarray}

Introducing an usual tortoise coordinate
\begin{eqnarray}
dr_{\star}={r^2+a^2\over \Delta}dr,
\end{eqnarray}
and resolving the equation (\ref{RR1}) in the form
\begin{eqnarray}
&&R_{s}={{\Delta^{-s/2}}\over \sqrt{r^2+a^2}}(\lambda^2+\mu^2
r^2)^{1/4}e^{-{i s}\arctan{( {\mu r\over \lambda}})}  \Psi_{s},
\end{eqnarray}
we have
\begin{eqnarray}\label{Psi1}
{d^2\Psi_{s}\over {dr_{\star}}^2}+\left[{dH_{s}\over
dr_{\star}}-H_{s}^2+{\Delta\over {(r^2+a^2)^2}}P_{s}
\right]\Psi_{s}=0,
\end{eqnarray}
where
\begin{eqnarray}
&&H_{s}=-{s\over {2(r^2+a^2)}}{d\Delta\over dr}-{\Delta\over
(r^2+a^2)^{3/2}}{d\sqrt{r^2+a^2}\over dr}-{s\Delta\over
{(r^2+a^2)}}{i \mu\over {\lambda-2 i s \mu r}}.
\end{eqnarray}
We will use Eq. (\ref{Psi1}) to study the late-time evolution of
the charged massive Dirac field.

\section{Late-time tail of the charged massive Dirac field}

The time evolution of a charged massive Dirac field in Kerr-Newman
spacetime is given by
\begin{eqnarray}
&&\Psi_{s}(r_{\star},t)=\int[G_{s}(r_{\star},r'_{\star};t)
\partial_{t}\Psi_{s}(r'_{\star},0)
+\partial_{t}G_{s}(r_{\star},r'_{\star};t)\Psi_{s}(r'_{\star},0)
]dr'_{\star},
\end{eqnarray}
where the black-hole (retarded) Green's function
$G_{s}(r_{\star},r'_{\star};t)$ is defined by
\begin{eqnarray}
&&\left\{{\partial^2\over {\partial r_{\star}}^2}-{\partial^2\over
{\partial t}^2}+{dH_{s}\over dr_{\star}}-H_{s}^2+\left[{-2i
ma\over {\varrho}^2}+{2s(r-M)\over {\varrho}^2}-{{4 sr\Delta}\over
{\varrho^4}}-{ i \mu\over {\lambda+2 i s \mu r}}{\Delta \over
{\varrho^2}}\right]{\partial\over {\partial t}}\right.\nonumber
\\
&&+\left.{{{m^2}{a^2}+2 i s(r-M)ma}\over
{\varrho^4}}-{\Delta(\mu^2 r^2+\lambda^2-(s+1/2)+2 i s e Q)\over
{\varrho^4}}\right.\nonumber
\\&& \left. +{{\Delta\over {\varrho^4}}{i\mu
(r-M)(s+1/2)+\mu ma\over {\lambda-2 i s \mu r}}}
\right\}G_{s}(r_{\star,},r'_{\star,};t)
=\delta(t)\delta(r_{\star}-r'_{\star}),
\end{eqnarray}
with $\varrho=\sqrt{r^2+a^2}$. The causality condition gives us
the initial condition $G_{s}(r_{\star},r'_{\star};t)=0$ for
$t\leq0$. In order to get $G_{s}(r_{\star},r'_{\star};t)$ we
calculate $G_{s}(r_{\star},r'_{\star};t)$ through the Fourier
transform
\begin{eqnarray} \label{Four1}
G_{s}(r_{\star},r'_{\star};t)={1\over
2\pi}\int_{-\infty+ic}^{\infty+ic}\tilde{G}_{s}(r_{\star},r'_{\star};\omega)e^{-i\omega
t}d\omega,
\end{eqnarray}
where $c$ is a positive constant. The corresponding inversion
formula is given by
\begin{eqnarray}
\tilde{G}_{s}(r_{\star},r'_{\star};\omega)=\int_{0^{-}}^{\infty}G_{s}(r_{\star},r'_{\star};t)e^{i\omega
t}dt.
\end{eqnarray}
The Fourier transform is analytic in the upper half $\omega$-plane
and it satisfies the equation
\begin{eqnarray}\label{GGG1}
&&\left\{{d^2\over dr_{\star}^2}+{dH_{s}\over dr_{\star}}-H_{s}^2
+{\Delta\over (r^2+a^2)^2}P_{s}
\right\}\tilde{G}_{s}(r_{\star},r'_{\star};\omega)=\delta(r_{\star}-r'_{\star}).
\end{eqnarray}

We define auxiliary functions
${\tilde{\Psi}_{\pm1}}({r_{\star},\omega})$ and
${\tilde{\Psi}_{\pm2}}({r_{\star},\omega})$ which are (linearly
independent) solutions to Eq. (\ref{Psi1}). Using the solutions
${\tilde{\Psi}_{\pm1}}$ and ${\tilde{\Psi}_{\pm2}}$, the
black-hole Greens's function can be constructed as
\begin{eqnarray}
\label{18}\ \ \ \
\tilde{G}_{\pm}(r_{\star},r'_{\star};\omega)=-\frac{1}
{W_{\pm}(\omega)} \left\{ \begin{array}{l}
\tilde{\Psi}_{\pm1}(r_{\star},\omega)\tilde{\Psi}_{\pm2}(r'_{\star},
\omega),\ \ r_{\star}<r'_{\star}, \\
\tilde{\Psi}_{\pm1}(r'_{\star},\omega)\tilde{\Psi}_{\pm2}(r_{\star},\omega),
\ \ r_{\star}>r'_{\star}, \end{array} \right.
\end{eqnarray}
where
$W_{\pm}(\omega)=W_{\pm}({\tilde{\Psi}_{\pm1}},{\tilde{\Psi}_{\pm2}})=
{\tilde{\Psi}_{\pm}}{\tilde{\Psi}_{\pm2,x}}-
{\tilde{\Psi}_{\pm2}}{\tilde{\Psi}_{\pm1,x}}$ is the Wronskian.

Morse and Feshbach  \cite{19} shown that massive tails exist even in
a flat spacetime due to the fact that different frequencies forming
a massive wave packet have different phase velocities. We will see
that at intermediate times the backscattering from asymptotically
far regions is negligible compared to the flat spacetime massive
tails that appear here. Hod, Piran, and Leaver \cite{13, 17} argued
that the asymptotic massive tail is associated with the existence of
a branch cut (in $\tilde{\Psi}_{2}$) placed along the interval
$-\mu\leq \omega\leq\mu$. This tail arises from the integral of the
Green function $\tilde{G}(r_*, r_*';\omega)$ around the branch
(denoted by $G^C(r_*, r_*';\omega)$) which gives rise to an inverse
power-law behavior of the Dirac fields. Therefore our goal is to
carry out $G^C(r_*, r_*';\omega)$.

When both the observer and the initial data are situated far away
from the black hole, (a large $r$ or equivalently, a low
$\omega$), we expand the wave equation (\ref{Psi1}) for the
charged massive Dirac field as a power series in $M/ r$ and $Q/
r$, neglecting terms of order $O((\omega/r)^2)$ and higher, we
obtain
\begin{eqnarray}\label{xi1}
&&\left\{{d^2\over {dr^2}}+\omega^2-\mu^2+{{4M \omega^2-2M \mu^2-2
e Q \omega}\over r} -{(\lambda/ \mu+2i s M+2am+8 M e Q)\omega\over
r^2}\right. \nonumber \\
&&\left. -{\lambda^2-2 i s e Q- Q^2(e^2+\mu^2)+(4M^2-a^2)\mu^2
\over r^2}\right \}\xi=0,
\end{eqnarray}
where $\xi=\sqrt{\Delta\over {r^2+a^2}}\Psi_{s}$. We take the transformation
\begin{eqnarray}
z&=&2\sqrt{\mu^2-\omega^2}r=2\varpi r,\nonumber \\
\xi&=&e^{-z/2}z^{1/2+b}\Phi,\nonumber \\
b^2&=&{1\over 4}+\lambda^2-2iseQ-Q^2(e^2+\mu^2)+(4M^2-a^2)\mu^2\nonumber \\
&&+({\lambda\over \mu}+2 i s M+2a m+8MeQ)\omega,\nonumber \\
k&=&{M \mu^2-eQ\omega\over \varpi}-2M \varpi,
\end{eqnarray}
then, Eqs. (\ref{xi1}) becomes the confluent hypergeometric equation \cite{hy}
\begin{eqnarray}
z{d^2\Phi\over dz^2}+(1+2b-z){d\Phi\over dz}-({1\over
2}+b-k)\Phi=0.
\end{eqnarray}
The two basic solutions required in order to build the Green's
function can be expressed as
\begin{eqnarray}
&&{\tilde{\Psi}_{1}}=A e^{-\varpi r}(2 \varpi r)^{1/2+b}M({{1\over
2}+b-k},{1+2b},{2\varpi r}),\\
&&{\tilde{\Psi}_{2}}=B e^{-\varpi r}(2 \varpi r)^{1/2+b}U({{1\over
2}+b-k},{1+2b},{2\varpi r}),
\end{eqnarray}
where $A$ and $B$ are normalization constants. The functions
$M(\tilde{k},\tilde{b},z)$ and $U(\tilde{k},\tilde{b},z)$
represent the two standard solutions to the confluent
hypergeometric equation \cite{hy}. $U(\tilde{k},\tilde{b},z)$ is a
many-valued function, so there is a cut in ${\tilde{\Psi}}_{2}$.
Then, we find that the branch cut which contribute to the Green's
function can be expressed as
\begin{eqnarray}
G^{C}(r_{\star},r'_{\star};t)&=&{1\over
2\pi}\int_{-\mu}^{\mu}\left[{{\tilde{\Psi}_{1}}(r'_{\star},\omega
e^{i \pi}){\tilde {\Psi}_{1}}(r_{\star},\omega e^{i \pi})\over
{W(\omega e^{i \pi})}}-{{\tilde{\Psi}_{1}}(r'_{\star},\omega
){\tilde {\Psi}_{1}}(r_{\star},\omega )\over
{W(\omega)}}\right]e^{-i \omega
t}d\omega \nonumber \\
&\equiv &{1\over 2\pi}\int_{-\mu}^{\mu}F(\varpi)e^{-i \omega
t}d\omega.
\end{eqnarray}
 We can get the following relations with the use of the
 Eqs. (13.1.32), (13.1.33),  and
(13.1.34) of Ref. \cite{hy}
\begin{eqnarray}
{\tilde{\Psi}}_{1}(2\varpi r)&=&A e^{-\varpi r}(2\varpi
r)^{{1\over
2}+b}M({1\over 2}+b-k,1+2b,2\varpi r), \\
{\tilde{\Psi}}_{2}(2\varpi r)&=&B{{\Gamma(-2b)}\over \Gamma
({1\over 2}-b-k)}e^{-\varpi r}(2\varpi r)^{{1\over 2}+b}M({1\over
2}+b-k,1+2b,2\varpi r)\nonumber\\&&+B{{\Gamma(2b)}\over \Gamma
({1\over 2}+b-k)}e^{-\varpi r}(2\varpi r)^{{1\over 2}-b}M({1\over
2}-b-k,1-2b,2\varpi r).
\end{eqnarray}
Using Eq. (13.1.22) of Ref. \cite{hy}, we get
\begin{eqnarray}
W(\varpi e^{i \pi})=-W(\varpi)=AB{{\Gamma(2b)}\over {\Gamma
({1\over 2}+b-k)}}4b \varpi.
\end{eqnarray}
Then, we find that $F(\varpi)$ is given by {\small
\begin{eqnarray}
&F(\varpi)&={{r_{\star}}^{{1\over 2}-b}{r'_{\star}}^{{1\over
2}+b}e^{-\varpi ({r_{\star}}+{r'_{\star}})}\over
{2b}}\left[M({1\over 2}+b-k,1+2b,2\varpi r'_{\star})M({1\over
2}-b-k,1-2b,2\varpi r_{\star})\right.\nonumber
\\&&-\left.M({1\over 2}+b+k,1+2b,2\varpi r'_{\star})M({1\over
2}-b+k,1-2b,2\varpi r_{\star} )\right]+{\Gamma (-2b)\Gamma
({1\over 2}+b-k)\over {\Gamma (2b)\Gamma ({1\over
2}-b-k)}}\nonumber
\\&& \frac{{e^{-\varpi ({r_{\star}}+{r'_{\star}})}}(4
\varpi^2 {r_{\star}{r'_{\star}}})^{{1\over 2}+b}}{4b \varpi}\left[M({1\over
2}+b-k,1+2b,2\varpi r'_{\star})M({1\over 2}+b-k,1+2b,2\varpi
r_{\star})\right. \nonumber
\\&&+\left.e^{(1+2b)i\pi}M({1\over
2}+b+k,1+2b,2\varpi r'_{\star})M({1\over 2}+b+k,1+2b,2\varpi
r_{\star})\right].
\end{eqnarray}}
In what follows, we will evaluate $G^{c}(r_*,r_*';t)$  at the
intermediate and very late times, respectively.

\subsection{Intermediate behavior of the charged massive Dirac field }

We First focus our attention on the intermediate behavior of the
charged massive Dirac field. When the late-time behaviors of
$G^{C}$ at the time scale $M\ll r \ll t \ll {M\over {(M \mu)^2}}$,
which means $k\ll 1 $, following Hod and Prian \cite{eg.3}
\cite{eg.4}, the effective contribution to the integral in Eq.
(\ref{Four1}) is limited to the range $\varpi=O(\sqrt{\mu/t})$.
This is due to the rapidly oscillating term $e^{-i \omega t}$,
which leads to a mutual cancellation between the positive and the
negative parts of the integrand. The condition $k\ll 1$ shows that
the $1/r$ term which describes the effect of backscattering off
the spacetime curvature from the asymptotically far regions is
negligible.

For the case $k\ll 1 $, noting that
$M(\tilde{k},\tilde{b},z)\simeq 1 $ as $z\rightarrow 0$, we get
\begin{eqnarray}
F(\varpi)&\simeq &-{{1+e^{(1+2b)i \pi}}\over 4\varpi b}{{{{\Gamma
(-2b)}{\Gamma ({1\over 2}+b)}}}\over {{{\Gamma (2b)}{\Gamma
({1\over
2}-b)}}}}{(2\varpi)^{1+2b}}{({r_{\star}}{r'_{\star}})^{{1\over
2}+b}}\nonumber \\
&=&{\pi\over sin(\pi b)}{{1+e^{(1+2b)i \pi}}\over
{{2^{1+2b}}b^2}}{{\varpi ^{2b}}\over \Gamma
(b)^2}{({r_{\star}}{r'_{\star}})^{{1\over 2}+b}}.
\end{eqnarray}
After tedious calculation, we get
\begin{eqnarray} \label{Green}
G^{C}(r_{\star},r'_{\star};t)={1\over 4}\int_{-\mu}^{\mu}{1\over sin(\pi b)}{{1+e^{(1+2b)i
\pi}}\over {{2^{2b}}{b^2}}}{{({r_{\star}}{r'_{\star}})^{{1\over
2}+b}}\over{\Gamma (b)^2}}{(\mu^2-\omega^2)^b}{e^{-i \omega
t}}d\omega.
\end{eqnarray}

\begin{figure}
\includegraphics[width=16cm]{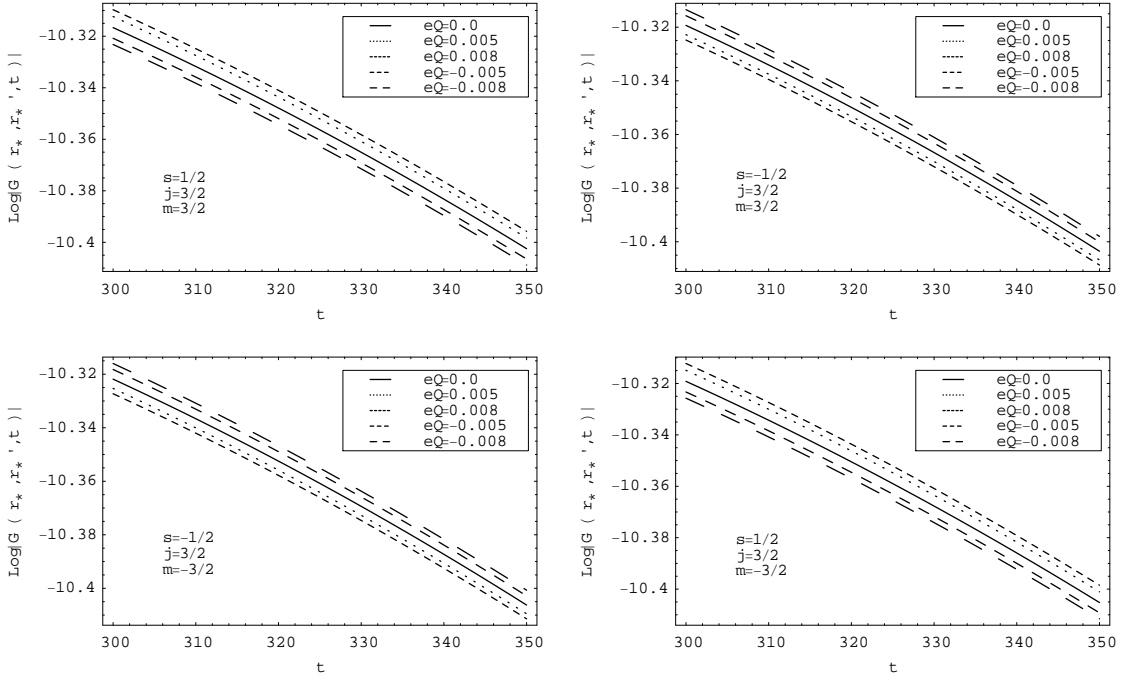}
\caption{\label{fig1} Graphs of ln$
|{G^{C}}({r_{\star},{r'_{\star}};t})|$ versus $t$ for different
$eQ$ with $\mu=0.01$, $a=0.05$ and  $\lambda > 0$. For $seQ>0$,
the larger the parameter $Q$ is, the more slowly the perturbation
decays. But for  $seQ<0$, the larger the parameter $Q$ is, the
more quickly the perturbation decays.} \end{figure}

\begin{figure}
\includegraphics[width=16cm]{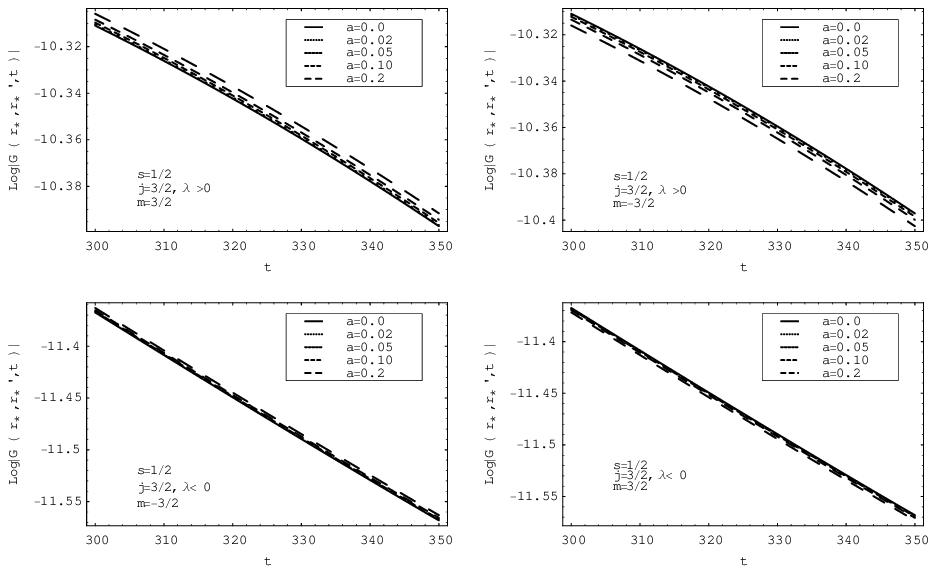}
\caption{\label{fig2}Graphs of ln$
|{G^{C}}({r_{\star},{r'_{\star}};t})|$ versus $t$ for $j={3/2}$,
with $Q=0.8$, $\mu=0.01$.  The rotating parameter $a$ speeds the
decay up for $m\lambda<0$ but slows it down for $m\lambda>0$. }
\end{figure}

\begin{figure}
\includegraphics[width=15cm]{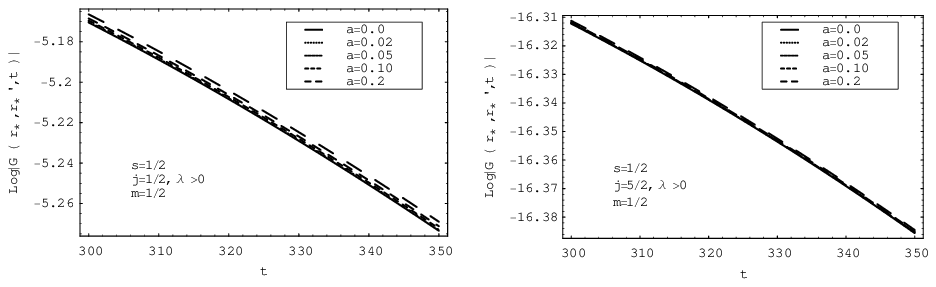}
\caption{\label{fig3}Graphs of ln$
|{G^{C}}({r_{\star},{r'_{\star}};t})|$ versus $t$ for $j=1/2$ and
$j=5/2$, with $Q=0.8$, $\mu=0.01$.  The rotating parameter $a$
speeds the decay up for $m\lambda<0$ but slows it down for
$m\lambda>0$. } \end{figure}

\begin{figure}
\includegraphics[width=16cm]{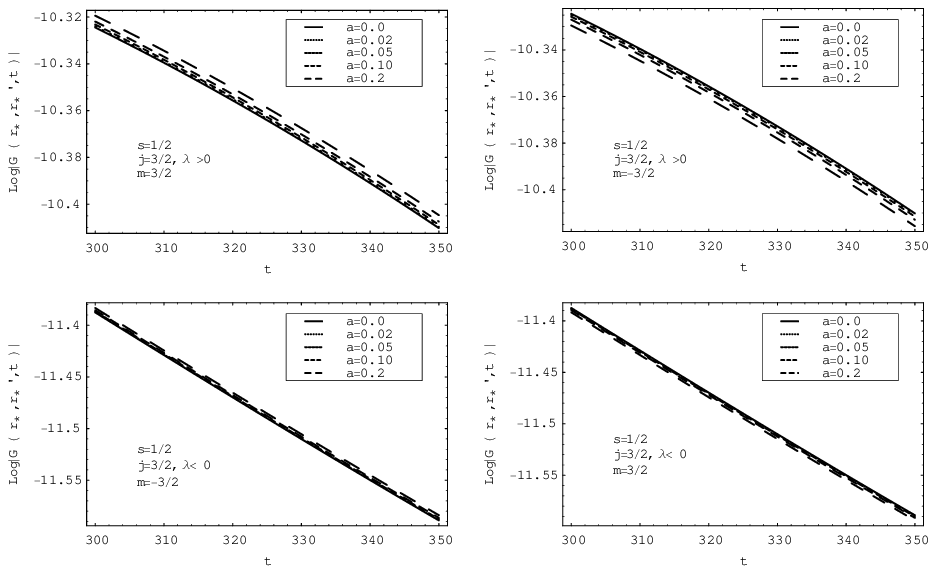}
\caption{\label{fig4} Graphs of ln$
|{G^{C}}({r_{\star},{r'_{\star}};t})|$ versus $t$, with $Q=-0.8$,
$\mu=0.01$, $j=3/2$.  The rotating parameter $a$ speeds the decay
up for $m\lambda<0$ but slows it down for $m\lambda>0$. }
\end{figure}

We work out the integral numerically and present the results in
the figures \ref{fig1}-\ref{fig4}. Figure \ref{fig1} describes ln$
|{G^{C}}({r_{\star},{r'_{\star}};t})|$ versus $t$ for different
$eQ$, and  $\lambda > 0$. We find that the dumping exponent
depends on the product $seQ$ of the spin weight, the charge of the
Dirac particles and the charge of the black hole, and $seQ>0$
slows the perturbation decay down but $seQ<0$ speeds it up. It is
the same as $\lambda < 0$ case. Figure \ref{fig2}, \ref{fig3}
illustrate ln$ |{G^{C}}({r_{\star},{r'_{\star}};t})|$ versus $t$
for different $a$ with $Q=0.8$ and $\mu=0.01$ and Figure
\ref{fig4} gives ln$ |{G^{C}}({r_{\star},{r'_{\star}};t})|$ versus
$t$ for different $a$ with $Q=-0.8$ and $\mu=0.01$. Figs.
\ref{fig2}, \ref{fig3} and \ref{fig4} tell us that the dumping
exponent depends on the quantum number $m$, the separation
constant $\lambda$ and the rotating parameter $a$, and also show
that, for both positive and negative electric charge of the black
hole, the rotating parameter $a$ slows the decay rate down for $m
\lambda > 0$ but speeds it up for $m \lambda < 0$.

\begin{figure}
\includegraphics[width=15cm]{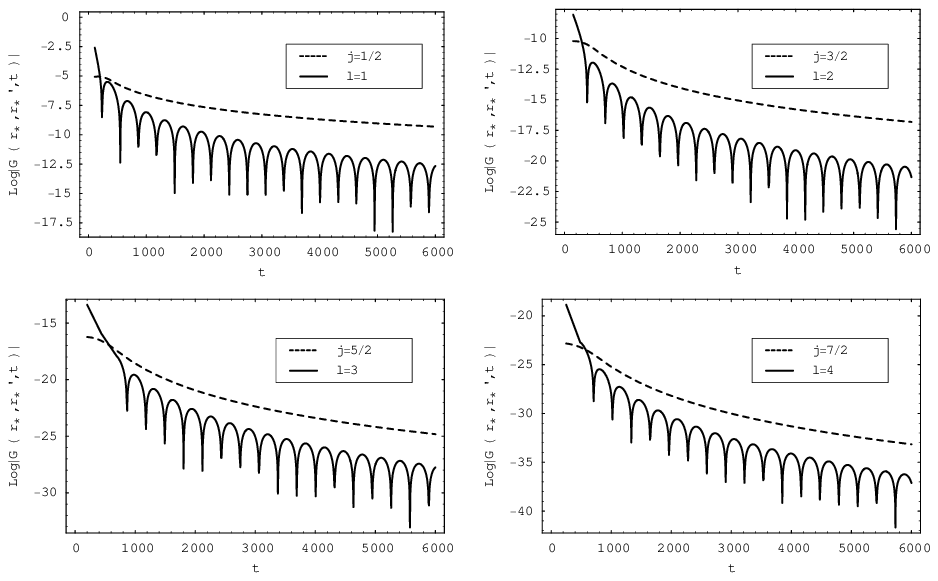}
\caption{\label{fig5} Graphs of ln$ |{G^{C}}({r_{\star},
{r'_{\star}};t})|$ versus $t$ with $\mu=0.01$, $Q=0.8$, $a=0.02$.
The dashed lines represent the result of the Green function of the
massive Dirac field with different $j$. For comparing, we also
show the corresponding result of the scalar field with real lines.
It is shown that the late-time behavior of massive Dirac fields is
dominated by a decaying tail without any oscillation, and the
decay of the massive Dirac field is slower than that of the
massive scalar field.} \end{figure}

To compare with the oscillating late-time behavior of the massive
scalar field in the Kerr-Newman background \cite{scalar}, we draw
the graphs of ln$ |{G^{C}}({r_{\star}, {r'_{\star}};t})|$ versus
$t$ with $\mu=0.01$, $Q=0.8$, $a=0.02$ in Fig \ref{fig5}. It is
shown that the  intermediate late-time behavior of charged massive
Dirac field is dominated by a decaying tail without any
oscillation, and the decay of the massive Dirac field which is
affected by the black-hole and field parameters is slower than
that of the massive scalar field.

\subsection{Asymptotically late-time tail}

The intermediate tail is not a final pattern of the perturbation
decay. At very late times, $\mu t \gg {1/(\mu M)^2}$, we find
\begin{eqnarray}
k\simeq {{M \mu^2-e Q \mu \over \varpi} }\gg  1,
\end{eqnarray}
which  means that  the $1/r$ term gives the dominant contribution
to the decay. That is to say, the backscattering from the
asymptotically far regions is important. Using the Eq. (13.5.13)
of Ref. \cite{hy}, we obtain
\begin{eqnarray}\label{FFF}
F(\omega)\simeq {\Gamma(1+2b)\Gamma(1-2b)r_{\star}
r'_{\star}\over 2b} \left[J_{2b}(\sqrt{\alpha r'_{\star}})
J_{-2b}(\sqrt{\alpha r_{\star}})-I_{-2b}(\sqrt{\alpha
r_{\star}})I_{-2b}(\sqrt{\alpha r_{\star}})\right]+&&\nonumber \\
{\Gamma(1+2b)^2\Gamma(-2b)\Gamma({1\over 2}+b-a)r_{\star}
r'_{\star}\over 2b \Gamma(2b)\Gamma({1\over 2}-b-a)}\left[J_{2b}
(\sqrt{\alpha r'_{\star}}) J_{-2b}(\sqrt{\alpha
r_{\star}})+I_{-2b}(\sqrt{\alpha r_{\star}}) I_{-2b}(\sqrt{\alpha
r_{\star}})\right],\nonumber \\
\end{eqnarray}
where $\alpha=8M\mu^2-8eQ\mu$, and $I_{\pm2b}$ are modified Bessel functions.

\begin{figure}
\includegraphics[width=8cm]{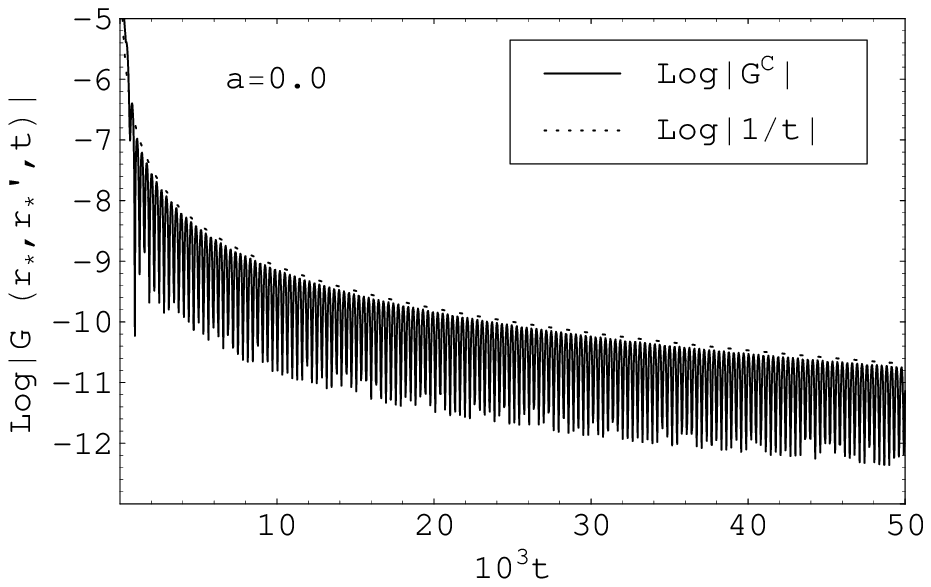}
\includegraphics[width=8cm]{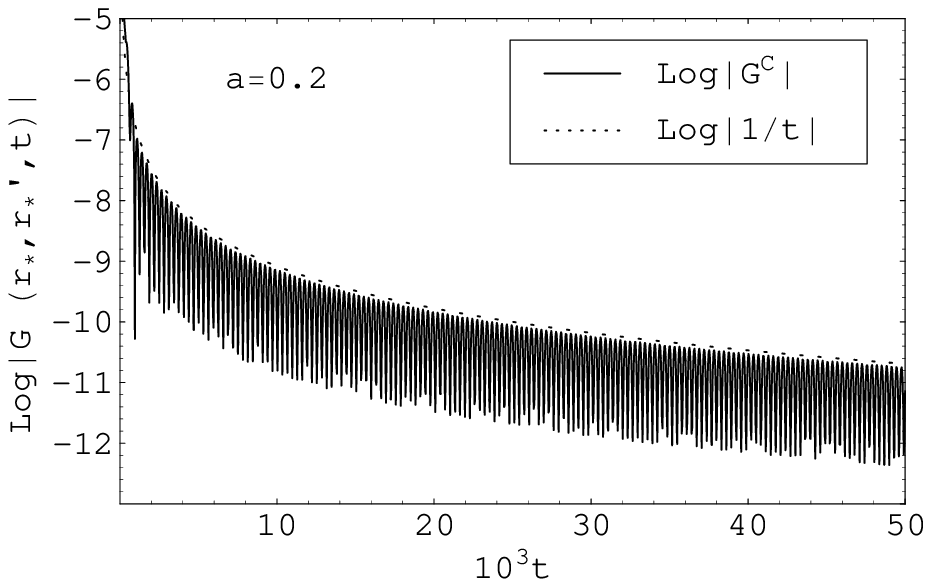}
\caption{\label{fig6} Graphs of ln$
|{G^{C}}({r_{\star},{r'_{\star}};t})|$ versus $t$ with $M=100$,
$\mu=0.01$, $Q=0.1M$, $e=0.002$, $s=-1/2$ and $j=1/2$ (the left
panel is for $a=0$ and right one for $a=0.2$) for the first term
(solid line).  The dashed line is $\sim \log
\left|\frac{1}{t}\right|$. From the figure we know, for different
$a$, that the late-time tail arising from the first term is $\sim
\frac{1}{t}$.} \end{figure}

It is easily to know that the factor $\Gamma(1+2
b)\Gamma(1-2b)/2b$ in the first term is not a constant. Therefore,
we use numerical method to study late-time behavior of the first
term and present the result in Fig. (\ref{fig6}). The figure
shows, for different value of $a$, that asymptotically late-time
tail arising from the first term is still $\sim t^{-1}$ although
the factor $\Gamma(1+2 b)\Gamma(1-2b)/2b$ is not a constant.

Then we consider the contribution of the second term to the decay
rate. Because
\begin{eqnarray}
{\Gamma(1+2b)\Gamma(1-2b)\over
2b}=-{\Gamma(1+2b)^2\Gamma(-2b)\over 2b\Gamma(2b)},
\end{eqnarray}
and the factor almost does not affect the asymptotically behavior
of the firs term, we define
\begin{eqnarray}
A={\Gamma(1+2b)^2\Gamma(-2b)r_{\star}r'_{\star}\over 2b \Gamma(2b) }
\left[J_{2b}(\sqrt{\alpha r'_{\star}})
J_{-2b}(\sqrt{\alpha r_{\star}})+I_{-2b}(\sqrt{\alpha r_{\star}})
I_{-2b}(\sqrt{\alpha r_{\star}})\right],
\end{eqnarray}
which approximates to a constant. Then, the second term in Eq.
(\ref{FFF}) can be rewritten as
\begin{eqnarray}\label{Four2}
G^{C}(r_{\star},r'_{\star};t) \simeq {A\over 2\pi}\int_{-\mu}^{\mu}
{\Gamma({1\over 2}+b-k)\over \Gamma({1\over 2}-b-k) }a^{-2b} e^{-i
\omega t }d\omega.
\end{eqnarray}
Using the saddle point integration method, the Eq. (\ref{Four2})
can be written as
\begin{eqnarray}\label{FFFF}
G^{C}(r_{\star},r'_{\star};t) \simeq {A\over 2\pi}\int_{-\mu}^{\mu}
e^{i(2\pi k-\omega t)}e^{i \phi} d\omega,
\end{eqnarray}
where the phase is defined by
\begin{eqnarray}
e^{i\phi}={{1+(-1)^{2b}e^{-2i\pi k}}\over {1+(-1)^{2b}e^{2i\pi
k}}},
\end{eqnarray}
and it  remains in the range $0 \leq \phi \leq 2\pi$, even if $k$
becomes very large.
\begin{figure}
\includegraphics[width=8cm]{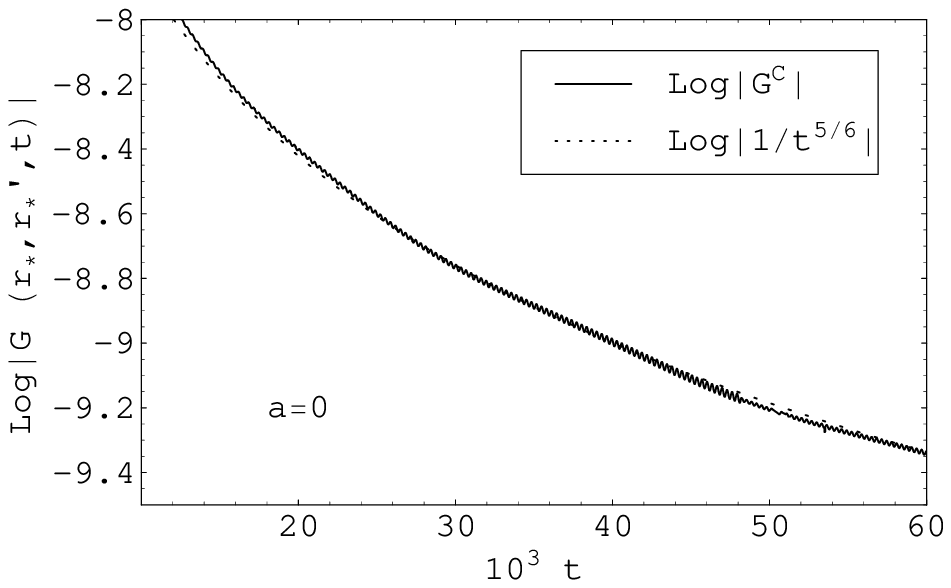}
\includegraphics[width=8cm]{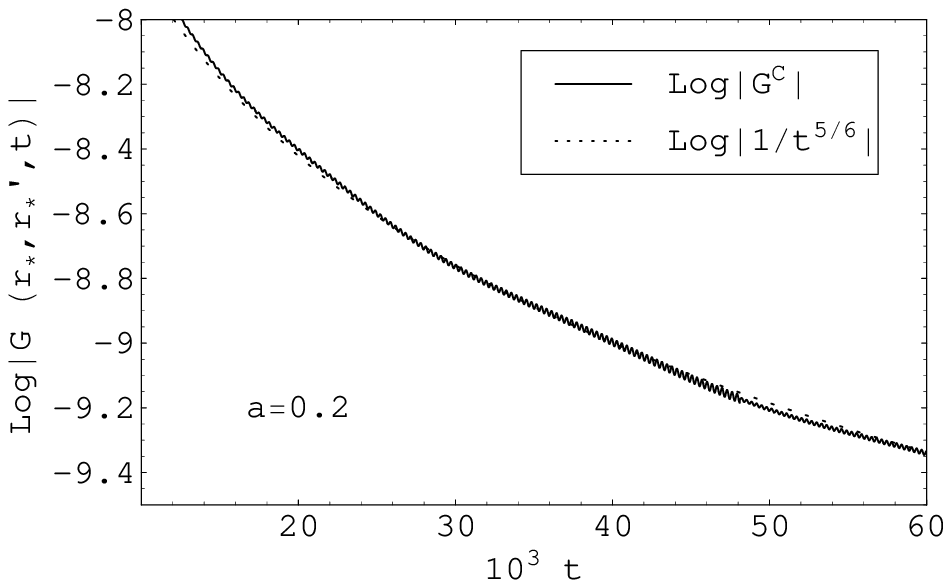}
\caption{\label{fig7} Graphs of ln$
|{G^{C}}({r_{\star},{r'_{\star}};t})|$ versus $t$ with $M=100$,
$\mu=0.01$, $Q=0.1M$, $e=0.002$, $s=-1/2$ and $j=1/2$ (the left
panel is for $a=0$ and right one for $a=0.2$) for the second term
(solid line).  The dashed line is $\sim \log
\left|\frac{1}{t^{5/6}}\right|$. The figure shows, for different
$a$, that the late-time tail arising from the second term is $\sim
\frac{1}{t^{5/6}}$.}
\end{figure}
 Solving
 \begin{eqnarray}
 {d\over d\omega}\ \ (2\pi k-\omega t)\ =\ 0,
 \end{eqnarray}
we get
\begin{eqnarray}
 \varpi_{0} \simeq \left[ 2\pi(M \mu^3-e Q \mu^2)\over t \right]^{1/3},
 \end{eqnarray}
at very large times (when $\mu t \gg {1/ (\mu M)^2}$ is
satisfied). In this region, most of the waves backscattered by the
space curvature are cancelled by each other, except for the
particular waves with the frequency $\omega_{0}\simeq\mu$.
Approximating the integral (\ref{FFFF}) in the immediate vicinity
of $\omega_{0}$, we find
\begin{eqnarray}
 G^{C}(r_{\star},r'_{\star};t)& \sim & {A \mu \over 2\sqrt{3}}
 (2\pi)^{5/6}(M \mu-e Q)^{1/3}(\mu t)^{-5/6} \nonumber \\
&&\times sin \left \{\mu t-[2\pi(M \mu -e Q)]^{2/3} (\mu
t)^{1/3}-\phi (\omega_{0}) -{\pi\over 4}\right\}.
 \end{eqnarray}
The equation shows that the decay rate of the asymptotic late-time
tail is exactly $t^{-5/6}$, and the oscillation has the period
$2\pi/\mu$ which is modulated by two types of long-term phase shift,
the one is $[2\pi(M \mu -eQ)]^{2/3} (\mu t)^{1/3}$ represents a
monotonously increasing phase  shift, the other is
$\phi(\omega_{0})$ represents a periodic phase shift.

In order to check the above analytical calculation, we also present
the numerical result of the second term in the Fig.~\ref{fig7} and
find that the decay rate of the asymptotically tail is  $\sim
t^{-5/6}$ for different value of $a$.

\section{Discussion and Summary}

The Teukolsky's master equations for the charged massive Dirac
fields in the Kerr-Newman spacetime are obtained by means of the
Newman-Penrose formulism. Then, both the intermediate late-time tail
and the asymptotic tail behavior of the charged massive Dirac fields
in the background of the Kerr-Newman black hole are studied by
considering the intermediate and very late-time solutions of the
Teukolsky's master equations. The results of the intermediate
late-time tail are presented by figures because we can not obtain
analytically Green's function $G^C(r_*, r_*'; t)$, as the parameter
$b$ in the integrand of the Green's function depends on the integral
variable $\omega$, especially for a factor $\lambda \omega/\mu$. We
learn from the figures that the intermediate late-time behavior of
charged massive Dirac field is dominated by a decaying tail without
any oscillation which is different from the oscillatory decaying
tails of the scalar field, and the decay of the massive Dirac field
is slower than that of the massive scalar field. We find that the
dumping exponent depends not only on the multiple number of the wave
mode but also on the product $seQ$ of the spin weight of the Dirac
field and the charges of the black hole and the fields ($seQ<0$
speeds up the decay of the charged massive Dirac fields but $seQ>0$
slows it down). We also find, for both positive and negative
electric charge of the black hole, that the rotating parameter $a$
slows the decay rate down for $m \lambda > 0$ but speeds it up for
$m \lambda < 0$.  And at very late time, the backscattering off the
curvature from the asymptotically far regions is dominant. We find,
for different value of the angular momentum per unit mass, that the
decay rate of the asymptotically late-time tail is  $t^{-5/6}$
through both numerical integral method and saddle-point integration
method, and the oscillation of the tail has the period of $2\pi /
\mu$ which is modulated by two types of long-term phase shifts. Our
result seems to suggest that the oscillatory $t^{-5/6}$ tail caused
by resonance back scattering at asymptotic late time may be a quit
general feature for the late-time decay of the charged massive Dirac
fields in stationary black-hole backgrounds.

\begin{acknowledgments}This work was supported by the
National Natural Science Foundation of China under Grant No.
10473004; the FANEDD under Grant No. 200317; the SRFDP under Grant
No. 20040542003; and the key project of the Hunan Provincial
Natural Science Foundation of China under Grant No.  05JJ0001.
\end{acknowledgments}

\appendix

\section{The separation constant}\label{app}

The angular equation (\ref{angular}) can be expressed as
\begin{eqnarray}\label{angular11}
\left\{{d^2  \over d\theta^2}+cot\theta{d  \over d\theta}+ {1\over
y}\left[{d^2y\over d\theta^2}+cot\theta {dy\over d\theta}+{a \mu
sin\theta\over -2s\lambda +\mu a cos\theta}{dy\over
d\theta}\right]+ \Theta_{s} +\lambda^2\right\}{_{s}\zeta_{jm}}=0,
\end{eqnarray}
where
\begin{eqnarray}
S_{s}&=&y {_{s}\zeta_{jm}},\nonumber \\
y&=&\sqrt{2s \lambda -a \mu cos\theta },\nonumber \\
\Theta_{s}&=&{\mu a sin\theta\over {\lambda-2s \mu a cos\theta}}H
- 2s{\partial H\over {\partial\theta}} - {1\over 2{sin^2\theta}}+
{\mu a cos\theta\over { 2(-2 s\lambda+a\mu cos\theta) }
}\nonumber\\&&-H^2 +{1\over 4}{cot^2\theta}- \mu^2a^2cos^2\theta.
\end{eqnarray}
For the slow rotating black hole,  ${_{s}\zeta_{jm}}$ and
$\lambda^2$ can be expanded as
\begin{eqnarray}
 {_{s}\zeta_{jm}}&=& _{s}P_{jm}+
{_{s}S_{jm}^{(1)}} a + {_{s}S_{jm}^{(2)}} a^2 + O
(a^3),\nonumber \\
\lambda^2&=& \lambda_{0}^2 + \lambda_{1}^2 a + \lambda_{2}^2
a^2+O(a^3),
\end{eqnarray}
where  $_{s}P_{jm}$ are the spherical harmonics with $s=\pm1/2$.
The functions  $_{s}P_{jm}$ and $_{s}\zeta_{jm}$
 satisfy the orthogonality relations
\begin{eqnarray}
\int{_{s}P_{j'm'}} {_{s}P_{jm}}d\Omega = \int
{_{s}\zeta_{j'm'}}{_{s}\zeta_{jm}}
d\Omega=\delta_{j'j}\delta_{m'm}.
\end{eqnarray}
Using the ordinary perturbation theory, we have
\begin{eqnarray}
&&[D_{0}+\lambda_{0}^2] {_{s}P_{jm}}=0\label{1},\\
&&[D_{0}+\lambda_{0}^2] {_{s}S_{jm}^{(1)}}+[D_{1}+\lambda_{1}^2] {_{s}P_{jm}}=0\label{2},\\
&&[D_{0}+\lambda_{0}^2] {_{s}S_{jm}^{(2)}}+[D_{1}+\lambda_{1}^2]
{_{s}S_{jm}^{(1)}} +[D_{2}+\lambda_{2}^2] {_{s}P_{jm}}=0\label{3},
\end{eqnarray}
with
\begin{eqnarray}
D_{0}&=&{d^{2}\over {d\theta^{2}}}+cot\theta{d\over
{d\theta}}-{2ms cos\theta \over {sin^2\theta}} -{1\over 2
sin^2\theta} -{m^2\over sin^2\theta}+{1\over
4}cot^2\theta,\nonumber\\
D_{1} &=& \left({ \mu cos\theta\over 4s \lambda_{0}}-{ \mu
m\over \lambda_{0}}+2m  \omega-2s \omega cos\theta \right),\nonumber\\
D_{2} &=& \left({m \mu \lambda_{1}^2\over 2\lambda_{0}^3}-{3 \mu^2
\over 16 s^2 \lambda_{0}^2}+{\mu \omega \over
\lambda_{0}}-\omega^2\right)- \mu \left({\lambda_{1}^2\over 8s
\lambda_{0}^3}+{2sm \mu \over \lambda_{0}^2} \right)cos\theta
\nonumber\\
&&+ \left({5\mu^2\over 16 s^2 \lambda_{0}}-{\mu \omega\over
\lambda_{0}}+\omega^2-\mu^2 \right)cos^2\theta.
\end{eqnarray}

From the lowest order equation (\ref{1}),  we have
\begin{eqnarray}\label{Lambda0}
\lambda_{0}^2=(j+{1\over 2})^2.
\end{eqnarray}

Multiplying equation (\ref{2}) by $_{s}P_{jm}$ from the left side
and integrating it over $\theta$, we obtain
\begin{eqnarray}\label{Lambda1}
\lambda_{1}^2=-2m \omega +{\mu m\over \lambda_{0}} + \left({\mu
\over 4s\lambda_{0}}-2s\omega\right){ms \over j(j+1)} .
\end{eqnarray}

Setting
\begin{eqnarray}
 _{s}S_{jm}^{(1)}= \sum_{j'}c_{jm}^{j'}{_{s}P_{j'm}}\label{4},
\end{eqnarray}
inserting this into equation (\ref{2}), then multiplying it by
$_{s}P_{j'm}$ from the left side and integrating it over $\theta$,
we have
\begin{eqnarray}
c_{jm}^{j'}={A\over {j'(j'+1)-j(j+1)}} \int {_{s}P_{j'm}}
cos\theta {_{s}P_{jm}}d\Omega,
\end{eqnarray}
where $A={\mu \over 2s(2j+1)}-2s\omega$. It is non-zero only for
$j'=j\pm 1$, so we obtain
\begin{eqnarray}
c_{jm}^{j+1}&=&{A\over 2(j+1)^2}
{\sqrt{(j+m+1)(j-m+1)(j+s+1)(j-s+1)\over
(2j+1)(2j+3)}},\nonumber\\
c_{jm}^{j-1}&=&-{A\over 2j^2} {\sqrt{(j+m)(j-m)(j+s)(j-s)\over
(2j-1)(2j+1)}}.
\end{eqnarray}

In order to obtain $\lambda_{2}^{2}$,  we multiply equation
(\ref{3}) by $_{s}P_{jm}$  from the left side,  and integrate it
over $\theta$, then we get
\begin{eqnarray}\label{Lambda2}
\lambda_{2}^2=&&-\int{_{s}P_{jm}} D_{1}{_{s}S_{jm}^{(1)}}d\Omega-
\int{_{s}P_{jm}}D_{2}{_{s}P_{jm}}d\Omega \\
=&&-2(j+1)\left( c_{jm}^{j+1} \right)^2+ 2j\left(c_{jm}^{j-1}
\right)^2\nonumber\\ &&-\left({m \mu \lambda_{1}^2\over
2\lambda_{0}^3}-{3 \mu^2 \over 16s^2 \lambda_{0}^2}+{\mu \omega
\over \lambda_{0}}-\omega^2\right) -\left({\lambda_{1}^2\over 8s
\lambda_{0}^3}+{2sm \mu \over \lambda_{0}^2} \right){\mu ms \over
j(j+1)}\nonumber\\ && - \left({5\mu^2\over 16s^2 \lambda_{0}}
-{\mu \omega\over \lambda_{0}}+\omega^2-\mu^2 \right)\left[{1\over
3}+{2\over 3} {[3m^2-j(j+1)][3s^2-j(j+1)]\over j(j+1)(2j+3)(2j-1)}
\right].
\end{eqnarray}
The integrals in the above equations are given by
\begin{eqnarray}
\int{_{s}P_{j'm}}cos\theta {_{s}P_{jm}}d\Omega&=&\sqrt{2j+1 \over
2j'+1}<
j, m, 1, 0  | j', m > < j, -s, 1,  0 | j', -s >,\nonumber \\
\int{_{s}P_{j'm}}sin^2\theta {_{s}P_{jm}}d\Omega&=&{2\over
3}{\delta_{j'j}}-{2\over 3}{\sqrt{2j+1 \over 2j'+1}}< j,  m, 2,  0
| j', m > < j, -s, 2,  0 | j', -s>.\nonumber
\end{eqnarray}

\end{document}